# SMART: SEMANTICALLY MASHUP REST WEB SERVICES

Rima Kilany[1] and Maroun Chamoun[2]

[1]ESIB, Saint-Joseph University - USJ, Beirut, Lebanon
rima.kilany@usj.edu.lb
[2]ESIB, Saint-Joseph University - USJ, Beirut, Lebanon
maroun.chamoun@usj.edu.lb

## ABSTRACT

*A mashup is a combination of information from more than one source, mixed up in a way to create something new, or at least useful. Anyone can find mashups on the internet, but these are always specifically designed for a predefined purpose. To change that fact, we implemented a new platform we called the SMART platform. SMART enables the user to make his own choices as for the REST web services he needs to call in order to build an intelligent personalized mashup, from a Google-like simple search interface, without needing any programming skills. In order to achieve this goal, we defined an ontology that can hold REST web services descriptions. These descriptions encapsulate mainly, the input type needed for a service, its output type, and the kind of relation that ties the input to the output. Then, by matching the user input query keywords, with the REST web services definitions in our ontology, we can find registered services individuals in this ontology, and construct the raw REST query for each service found. The wrap up from the keywords, into semantic definitions, in order to find the matching service individual, then the wrap down from the semantic service description of the found individual, to the raw REST call, and finally the wrap up of the result again into semantic individuals, is done for two main purposes: the first to let the user use simple keywords in order to build complex mashups, and the second to benefit from the ontology's inference engine in a way, where services instances can be tied together into an intelligent mashup, simply by making each service output individuals, stand as the next service input.*

## KEYWORDS

*REST Web Services, Mashups, Ontology & Semantic web*

## 1. INTRODUCTION

Mashups are changing the way we think about applications, it is no more the OS, API, Application paradigm, it is now the internet, web services, mashups stack. With more than 10 new mashups per day, mashups have become an important feature of the internet, and have evolved from ready-to-use mashups, like the one we can find on Programmable Web today [1] to the ready-to-build mashups using tools like yahoo pipes [2]. Being mainly targeted to work with RSS feeds (item lists), the DERI pipes project [3] comes up with a pipes design, conceptual model and implementation that specifically targets graph based RDF data and allows the developer to quickly prototype (semantic) Web applications using RDF [4]. This project fills the need that comes from the fact that an increasing amount of RDF data is becoming available, from widely used applications such as DBLP [5], DBpedia [6], blogs, wikis, forums, etc. that expose their content in different RDF-based formats such as SIOC [7] or FOAF [8], in the form of RDF/XML, or RDF statements embedded or extractable from HTML/XML pages by technologies such as GRDDL [9] or RDFa [10].

At the level of web services description, OWL-S is an ontology, within the OWL-based framework of the Semantic Web, for describing Semantic SOAP Web Services [11]. It enables users and software agents to automatically discover, invoke, compose, and monitor Web resources offering services, under specified constraints. Although having substantial differences with OWL-S, WSMO [15] is another major effort with the same goal.

WSMO-Lite [18] and MicroWSMO [14] both evolved from the WSMO framework. While WSMO-Lite uses SAWSDL [16] to annotate WSDL-based services, MicroWSMO uses the hRESTS microformat [17] to annotate RESTful APIs and services. Both frameworks share an ontology for service semantics.

Knowing that REST web services are now mostly used on the web, the SMART framework we expose in this article defines an ontology for describing REST web services, in order to enable their automatic discovery, invocation, and composition. A simple web application with a Google- like search interface is implemented as a proof of concept. It shows how the user can mashup services he needs to call without any programming knowledge, at the click of a finger.

In fact, what does it take to describe a service semantically? The answer is simple: The application that needs to call a service (or the integration framework that mashes them up) should have automatic access to information which normally a developer seeks to know and acquire in order to be able to call that service: What does the service do, how to build up the request, and how to extract information from the response.

With SMART, the user can build up the mashup he wants, in order to get access to whatever information he needs, without searching for it on the internet, and combining numerous data to obtain the result.

The rest of this paper is organized as follows: Section 2 details the ontology we defined for describing REST Web services. Section 3 describes the SMART platform. Section 4 explains the interaction with the user via the Google-like simple search interface. The SMART web service execution details are given in section 5. Section 6 describes the steps to add a new SMART service to the ontology. Finally conclusions and future work are given in Section 0.

## 2. SMART: SERVICE ONTOLOGY DEFINITION

In this section, we begin by describing the logic behind the semantic description we gave to a REST web service, this is what we call the service abstract model, then we detail all the classes and properties we needed to define in order to complete this description. We used the Protégé 4.1.0 Ontology editor to accomplish this task [19], and named our owl ontology file "services.owl".

### 2.1. The service abstract model

A SMART service is an operation with inputs and outputs. At the lowest level (The REST architectural level), inputs are simply named parameters holding literal data. Outputs are JSON or XML documents containing the returned data in a well-defined structure. These low level parameters should be semantically described, in order to be able to automate the calls to the service, as well as the extraction of the results.

This is done by defining for each service, at the ontology level, a class that wraps up the REST inputs, and a class that wraps up the output: This is what we defined as the Logical input and output Parameters of a service. These logical parameters are defined to encapsulate the definition of the raw REST parameters, as well as any other logical child parameters. They are tree-like structures, where the root is a logical parameter, while the branches can also be logical, and the leaves are surely raw REST literal parameters (string, double, etc…).

For example, let us consider a REST service that waits for geographic coordinates as input, and returns the corresponding region (name and country). At the REST level, this service admits two decimal parameters "*lng*" for longitude and "*lat*" for latitude, while at the semantic level, this service should admit a logical entry encapsulating the geographic coordinates, and which will be represented by a class named Location.

The REST output of such a service shows a Region, with a name, and the country to which the Region belongs. **Table 1** shows the XML output and the corresponding Logical parameter Region:

Table 1 Service Output

| | |
|---|---|
| XML output | <resp><br> <name>beirut</name><br><country><br><name>Lebanon</name><br> <id>LB</id><br> </country><br> </resp> |
| Logical Parameter Region (Turtle notation) | :Beirut a :Region:<br>:name “Beirut”@en;<br>:inCountry [ a :Country;<br>:name “Lebanon”@en;<br>:id “LB”^^xsd:string.]; |

The logical parameter Region (defined as a class Region in the ontology), is tied to the REST parameter name of the region, and the details of the country are tied to another sub-logical parameter Country (another class of the ontology).

By following the same logic, we tied up the input to the output of the service (see Figure 1), so the platform can establish that the coordinates at the input are for the Beirut Region output, and they are logically related.

```
:Beirut   a             :Region:
          :name         “Beirut”@en;
          :inCountry    [    a            :Country;
                             :name        “Lebanon”@en;
                             :id          “LB”^^xsd:string.];
          :location     [    a            :Location;
                             :longitude   “35.49442”^^xsd:decimal;
                             :latitude    “33.88894”^^xsd:decimal.]
```

Figure 1 SMART Logical Input Output Relation

This is achieved simply by defining two kinds of properties at the ontology level:

a. Properties that tie the REST Inputs and outputs to the corresponding Logical Parameters, and
b. Input/Output Relation properties that tie an output to its corresponding input.

These properties and the whole structure of the ontology will be explained in the next sub-section.

## 2.2. The ontology structure

The SMART ontology is divided into two main parts (see Figure 2): a part dedicated to the concepts related to the description of a REST web service and having the parent class ServiceThing as the root of all its concepts, and a part dedicated to the entities manipulated by services (inputs and outputs of a service), and having the class DomainThing as the root class. So, the description of a service parameter would be in a class Parameter under the ServiceThing class, while the description of a book or a Person would be under the class DomainThing.

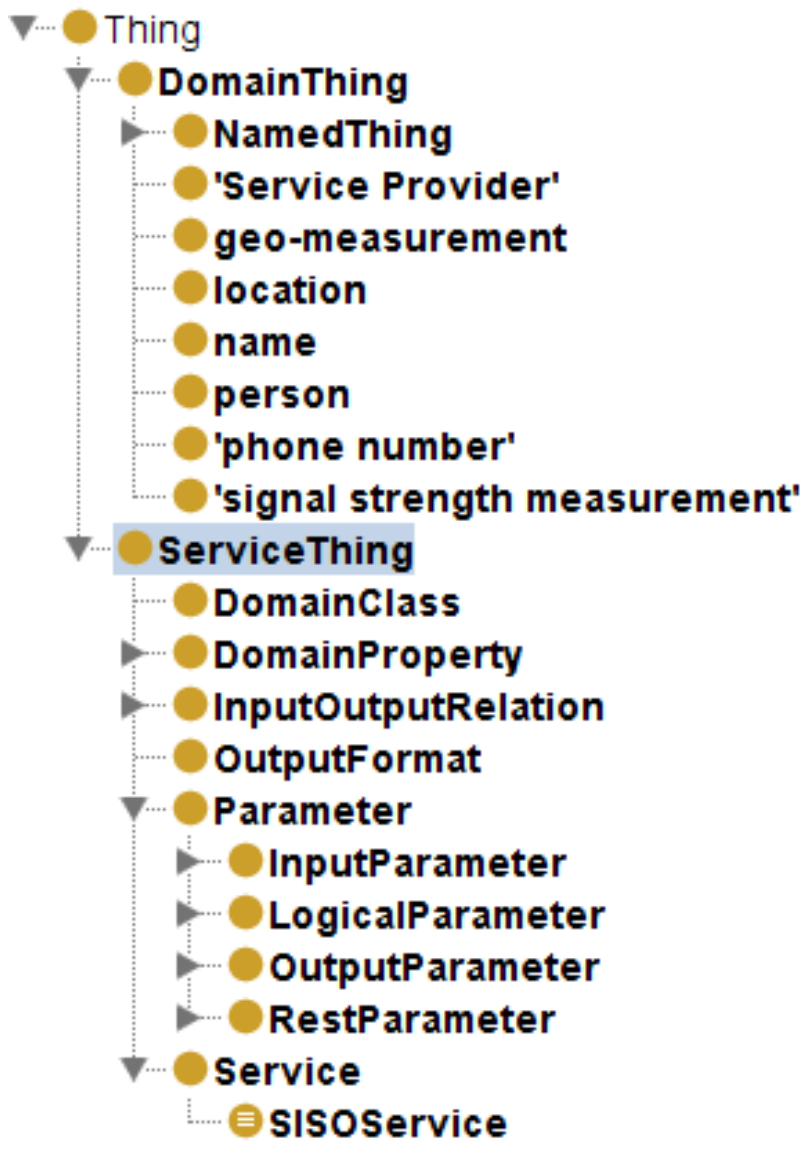

Figure 2 SMART Ontology classes

### 1) Main SMART Ontology classes

- Service: represents a RESTful web service.
- SISOService (Single Input, Single Output Service): services having exactly one logical Root input (RootLogicalInput) and one logical root output (RootLogicalOutput) (see Figure 3).

**Service** *and* (**hasRootInput** *exactly* 1 Parameter) and (**hasRootOutput** *exactly* 1 Parameter)

Figure 3 Single Input - Single Output definition

- Parameter: Superclass of all parameters.
- InputParameter: class representing entry parameters.
- OutputParameter: class representing output parameters.
- LogicalParameter: class representing logical parameters.
- LogicalInputParameter: class representing logical entry parameters.
- RootInputParameter: LogicalInputParameter(s) which are at the root.
- SubInputPrameter: LogicalInputParameter(s) which are not RootInputParameter(s).
- LogicalOutputParameter: class representing logical output parameters.
- RootOutputParameter: LogicalOutputParameter(s) which are at the root.

- SubOutputParameter : LogicalOutputParameter(s) which are not RootOutputParameter(s) (same relations as the input)
- RestParameter: class representing REST parameters. We should note that RestParameter class is disjoint with the LogicalParameter class.
- RestInputParameter: class representing REST entry parameters.
- StaticRestInputParameter: class representing constant REST entry parameters, e.g. a service key.
- VariableRestInputParameter:
  RestInputParameter(s) which are not StaticRestInputParameter(s).
- RestOutputParameter: class representing REST output parameters.
- InputOutputRelation: Input/Output Relations class, super class of:
  - InputToOutputRelation: class of InputOutputRelation(s) that ties an entry to an output.
  - OutputToInputRelation: class of InputOutputRelation(s) that ties an output to an entry.

**2) Main SMART Properties**

a) Object Properties:

The ontology Object Properties we defined are shown in Figure 4 and detailed here below:

- topDomainObjectProperty: super property of all properties joining DomainThing individuals: rdfs:domain and rdfs:range are of type DomainThing, as for example, for the ownerOf sub-property.
  We can be as specific as we need, by defining sub-properties, as for example hasPhoneNumber which is a sub-property of ownerOf: if a person has a phone number x, she is the owner of this phone number, but this can't be stated in the inverse way, simply because *owner of* is more generic than the hasPhoneNumber sub-property, which is defined with Person as rdfs:domain and PhoneNumber as rdfs:range. So when ownerOf is used, its meaning will depend on the classes it acts on, while the meaning of hasPhoneNumber can be directly deduced.
  We can also define equivalent properties like ownerOf which is equivalent to the owns and has properties, as well as inverse properties like ownedBy and of, in order to give meaning to these terms at the user level requests. The user will use the friendly labels we associated to these properties to build his request, and should not be aware of their exact definitions at the ontology level, while typing his request, as we will explain in Sections 4 and 5.

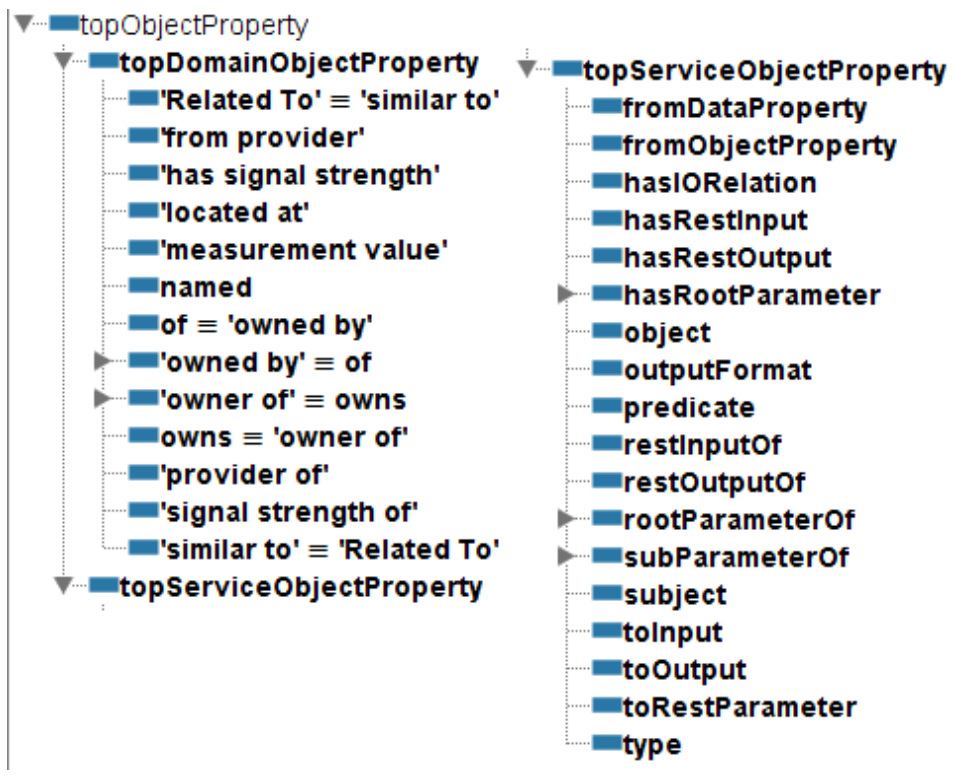

Figure 4 SMART object properties

- topServiceObjectProperty: super property of all properties joining ServiceThing individuals. Its sub-properties are:
  - fromDataProperty: associates to a VariableRestParameter a DomainDataProperty that ties a REST parameter of a service to the literal value given at the entry or read at the output.
  - fromObjectProperty: specifies the DomainObjectProperty that will be used to join a LogicalParameter (SubLogicalParameter more specifically) to its parent (see Figure 5).

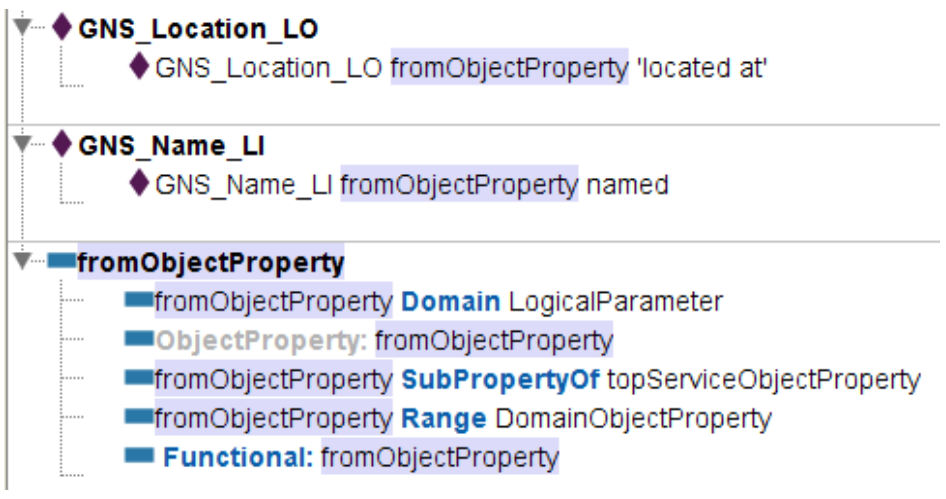

Figure 5 fromObjectProperty definition and usage examples

  - hasIORelation: associates an InputOutputRelation to a service individual. We can associate as many Input to Output relation individuals to a service individual as needed, but the value of this property can also be inferred by the following two SWRL Rules, so no need to specify it manually.
    - Rule 1: rootParameterOf(?rparam, ?service), subParameterOf(?sparam, ?rparam), subject(?rel, ?sparam) -> hasIORelation(?service, ?rel)
    - Rule 2: rootParameterOf(?rparam, ?service), subject(?rel, ?rparam) -> hasIORelation(?service, ?rel)

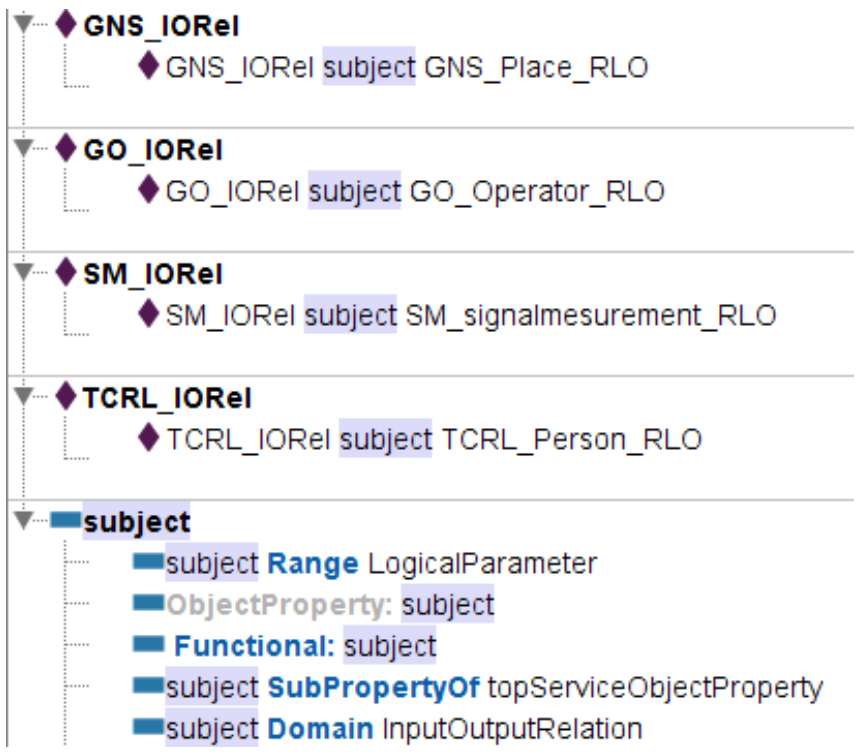

Figure 6 Input/Output Relations

We can see in Figure 6 that we can infer the hasIORelation property for each SMART REST service by using respectively the GNS_IORel, GO_IORel, SM_IORel expressions, combined with the Rule2 above.

  - hasRestInput: associates a RestInputParameter to a Service. Only REST parameters are explicitly tied to a service. Logical parameters are indirectly and implicitly tied to a service via the REST parameters that they are tied to.
  - hasRestOutput: associates a RestOutputParameter to a Service.
  - hasRootParameter: represents the implicit relation that ties a Service to its RootOuput Parameter(s) and RootInput Parameter(s).

- hasRootInput represents the implicit relation that ties a Service to its RootInput Parameter(s).
- hasRootOutput: represents the implicit relation that ties a Service to its RootOuput Parameter(s).
- hasRootInput and hasRootOutput properties can be specified manually or inferred from the Rule that ties a RestParameter to its LogicalInput at one hand, and to the Service at another hand (Cf. Rules: paragraph 5) of Sub-Section 5)).
- subject, predicate, object : Properties of an InputOutputRelation ; an InputOutputRelation declares that a LogicalParameter (defined by the subject property) should be tied to the another LogicalParameter (defined by the object property) through the DomainObjectProperty specified by the predicate property (see Figure 7 for an example).

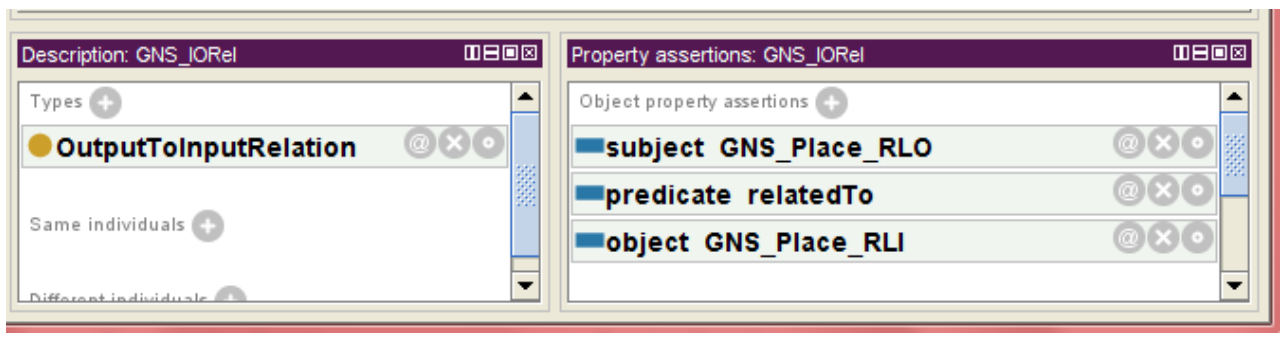

Figure 7 GeoNamesSearch Service IO Relation Individual

- restInputOf : inverse of hasRestInput
- restOutputOf : inverse of hasRestOutput
- rootParameterOf : inverse of hasRootParameter
- rootInputOf : inverse of hasRootInput

In order not to tie each rootInputParameter to the service it belongs to, the following SWRL rule RootInputParameter(?rootInput), hasRestInput(?service, ?restInput), subInputOf(?restInput, ?rootInput) -> rootInputOf(?rootInput, ?service). (for more detail, cf. Rules: paragraph 5) of Sub-Section 5)).

- rootOutputOf : inverse of hasRootOutput
- subInputOf: transitive property joining an InputParameter to his ancestor LogicalInputParameter(s) in the tree.
- subOutputOf: transitive property joining an OutputParameter to his ancestor LogicalOutputParameter(s) in the tree.
- subParameterOf: super property of subInputOf and subOutputOf.
- toInput: inverse of fromLogicalInput : ties a LogicalInputParameter to its InputParameter children in the tree.
- toOutput: inverse of fromLogicalOutput : ties a LogicalOutputParameter to its OutputParameter children.

By using the properties fromLogical{Input|Output} we can go up in the parameters tree, while to{Input|Output} will help us go down. The constructers of Logical{Input|Output} use to{Input|Output} respectively.

- toRestParameter : inverse of fromDataProperty
- type: property that specifies the type (DomainClass) of the data represented by a LogicalParameter.

b) Datatype Properties

- topDomainDataProperty: super property of all properties that tie DomainThing individuals to literal values. Figure 8 is an example of a topDomainDataProperty sub-property definition.

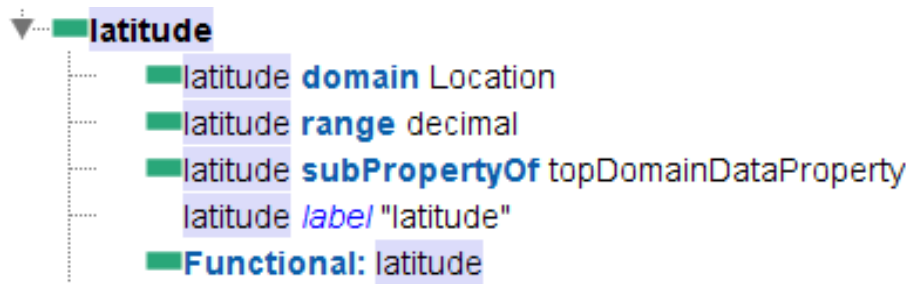

Figure 8 latitude property

- topServiceDataProperty: super property of all properties that tie ServiceThing individuals to literal values. We describe below all of its sub-properties:
  - endpoint: associates to a Service a base URL. REST parameters and their values will be concatenated to this URL in order to build the http GET request to the service.
  - mandatory: a property having a boolean value, in order to determine if a RestInputParameter is mandatory or not.
  - parameterValue: a property that gives to a StaticRestInputParameter its value.
  - parameterName: a property that gives a RestInputParameter its name.
  - resultXPath: a Service property showing the context of the results in the result XML document. If it is a simple structured document it could be simply '/'. In the case of a list structured document, it should point out to the location of a specific result node, e.g. in the case of the GeoNamesSearch resultXPath would be set to /geonames/geoname.
  - rootOutputXPath: property of a RootOutputParameter specifying the address of the XML node where to locate the root node of the output, relatively to the resultXPath of the service, e.g. it would be set to '.' for the GeoNamesSearch root logical output GNS_Place_RLO.
  - restOutputXPath: property of a RestOutputParameter specifying the address of the XML node holding the literal value relatively to the rootOutputXPath of the RootOutputParameter from which it descends, e.g. the GeoNamesSearch latitude REST Output GNS_lat_RO would be set to ‘lat’.

In conclusion, in this model, the Rest parameters are tied to the corresponding service by the property restInputOf. While the Logical parameters are not tied to the service directly, they are tied with each other and with the REST parameters. One should traverse the logical parameters tree, in order to get to the REST parameters in order to discover the service. In the coming version of SMART, we will tie the logical root parameters to the service, and deduce the remaining information about the service.

### 3) Service Individuals

We will see in details in Section 6, how to add a new SMART service. The SMART ontology already supports four different services that are fully described in the services.owl ontology file, by their corresponding individuals (see Figure 9).

- The GeoNames Search Service that describes the GeoNames geographical service: http://www.geonames.org/

• The GetOperator Service we defined in order to find the telecom operator for a certain mobile number.
• The signal measurement Service we defined in order to filter a signal strength measurement file (in csv format) for a certain provider.
• The TrueCallerReverseLookup that accesses the TrueCaller Service: http://www.truecaller.com/

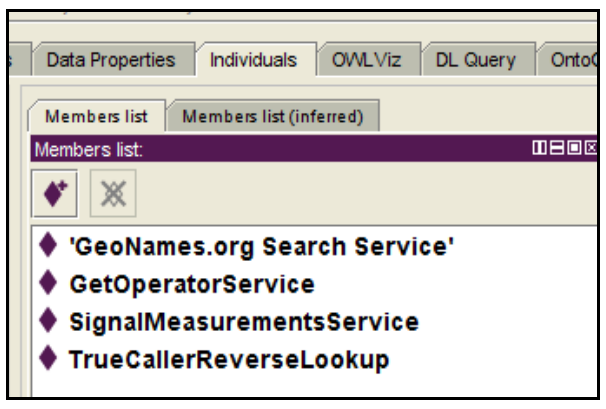

Figure 9 SISOService individuals

### 4) IORelations

Input Output Relational individuals are used in the SMART platform in order:

• To generate the output of a service in a way to show how the output individuals are tied to the input individuals.
• To be able to find a service by matching its IORel object property with available services, through the important information about the service (subject, predicate, object) that this property can provide us with. (Cf the matchService description in Section 5, where we will detail how to search for a service, through the ontology service individuals, by matching up some of their specific properties with the user query keywords).

In order to let the user use different keywords in his requests, we can define equivalent predicate IORelation properties, e.g. :similarTo owl:equivalentProperty :relatedTo (see Figure 4)

### 5) Rules

While we can express with OWL class expressions, that an individual belongs to a certain class, we are not able to deduce the value of a property respecting a number of conditions, e.g. if we know that x is the brother of y, and y is the father of z, we should then know that x is the uncle of z.

This is where we might use the OWL2 property chains [12] which have some limitations, or SWRL [13].

• The first SMART Rule shows how we can infer that a certain rootinput individual is associated to a service individual without the need to specify it explicitly for each service individual:

  **Rule 1:** RootInputParameter(?rootInput), hasRestInput(?service, ?restInput), subInputOf(?restInput, ?rootInput) -> rootInputOf(?rootInput, ?service)

  Rule 1 Explanation:

  - ?rootInput is a RootInputParameter,
  - it exists an instance of Service ?service having a RestInputParameter ?restParameter,
  - ?restParameter is a sub-parameter of ?rootInput

Then we can infer that ?rootInput is the rootInput of ?service

In fact, we have:

a. Defined, at the ontology level, that the fromLogicalInput is a subProperty of subInputOf: :fromLogicalInput rdfs:subPropertyOf :subInputOf

b. subInputOf is transitive this is the reason why an InputParameter (logical or REST) verifies this property as well as all his ancestors in the tree of the input parameters.

- The Second Rule helps us infer the IORelation individual associated to a certain Service individual, without the need to create it manually.
  Rule 2: rootParameterOf(?rparam, ?service), subject(?rel, ?rparam) -> hasIORelation(?service, ?rel)

  Rule 2 explanation in the case of the GeoNamesSearchService:

  As we can see:

  - GeoNamesSearch Service has GNS_Place_RLO for LogicalOuputParameter
  - GNS_IORel is the subject of GNS_Place_RLO

  Then we can say that the GeoNamesSearch Service hasIORelation GNS_IORel.

The rest of the rules are similar to those we have explained above.

**6) Labels**

An rdfs:label is associated to each topDomainObjectProperties. These labels are mandatory for the functioning of the requests; in fact the user will type: “find the phone number of this person” and in this case he will be using the label ‘phone number’ associated to the topDomainObjectProperty PhoneNumber.

**7) Punning**

The punning in OWL 2 has the objective to enable the use of classes and proprieties as subjects or objects of proprieties.

The following example (in turtle format) declares a class Person as the object of the type property:

:personParam :type :Person. This is made possible, by declaring the class Person as an individual of the owl:Thing class as follows: :Person a owl:Class, owl:Thing.

This is called the Class-Individual punnig.

The same thing can be done for the properties, as an example:

:nameparam :fromDataProperty :firstName

firstName is an owl:DataTypeProperty, and can be an object for the fromDataProperty property, if we declare :firsName as follows:

:firstName a owl:DatatypeProperty, owl:Thing;

rdfs:range xsd:string;

rdfs:domain :Person.

This is called the Property-individual punning.

In order to group our punned individuals according to their specific roles, we choose to define specific classes and not just use the owl:Thing class, which could also have been used:

- DomainClass: Used to do the class-individual punning, in the case of the DomainThing classes: The DomainClass individuals like Location, Name, Person will be used as objects of a property, i.e. GNS_Location_LO type Location, GNS_Location_LO fromObjectProperty located.
- DomainProperty: The super class of DomainObjectProperty and DomainDataProperty, which are used to do the Property-Individual punning respectively, in the case of the topDomainObjectProperty and topDomainDataProperty individuals so they can be used as objects of another property.

**8) Naming conventions**

The naming conventions we respected at the ontology level are the following:

- A service is named by the camel case convention (i.e.: GeoNamesSearch).
- The input/output parameters as well as the relation individuals names should be prefixed with an abbreviation of the service they are attached to (i.e. for the GeoNamesSearch the prefix would be GNS, and an input/output relation would be: GNS_IORel)
- A suffix indicates the type of the individual:
  - RI for a RestInputParameter
  - RO for a RestOutputParameter
  - LI for a LogicalInputParameter
  - LO for LogicalOutputParameter
  - RLI for a RootLogicalInputParameter
  - RLO for a RootLogicalOutputParameter
  - SI for a StaticRestInputParameter
  - IORel for an InputOutputRelation

As examples we give:

- GNS_Place_RLI, GNS_Place_RLO (the root input and respectively the root output of the GeoNamesSearch service)
- GNS_IORel (the InputOutputRelation instance which confirms that the GeoNamesSearch service input and output are tied by the relatedTo relation).

# 3. The SMART Platform

## 3.1. The SMART Architecture

At a high level, SMART defines an ontology of services, and an execution engine that interacts with the ontology in order to discover and execute the REST web services that the user needs to call. The execution results may be shown as a list, or on a map, in case the output encapsulates geographic location coordinates.

At the infrastructure level, SMART is mainly divided into two modules: the core module and the app module.

The core module contains all the generic classes that could be used by any external applications in order to interact with our ontology (i.e. and as a future work, a graphical mashup editor for REST web services with semantic support). It is itself divided into a utility package and a service package. It is the service package that contains all the classes that wrap up the definition of a Service, its parameters, as well as the classes that contain the service execution engine interacting with the defined ontology.

The app package is more specific to the prototype web application that waits for the request of the user in a sentence form, and is mainly dedicated to services having one logical input and one logical output, knowing that each logical entry can be formed of many other physical and/or logical entries (logical stands up for semantic).

The choices we made at all the levels of the SMART architecture, has been validated by the implementation of a web application we called "SMARTWeb" (cf. Section 4) and which we defined in the app.web package. This package encapsulates the classes forming the web interface which is the front-end part the user interacts with.

## 3.2. SMART Components interaction

The SMART main components and their interaction together are shown in the following diagram (Figure 10):

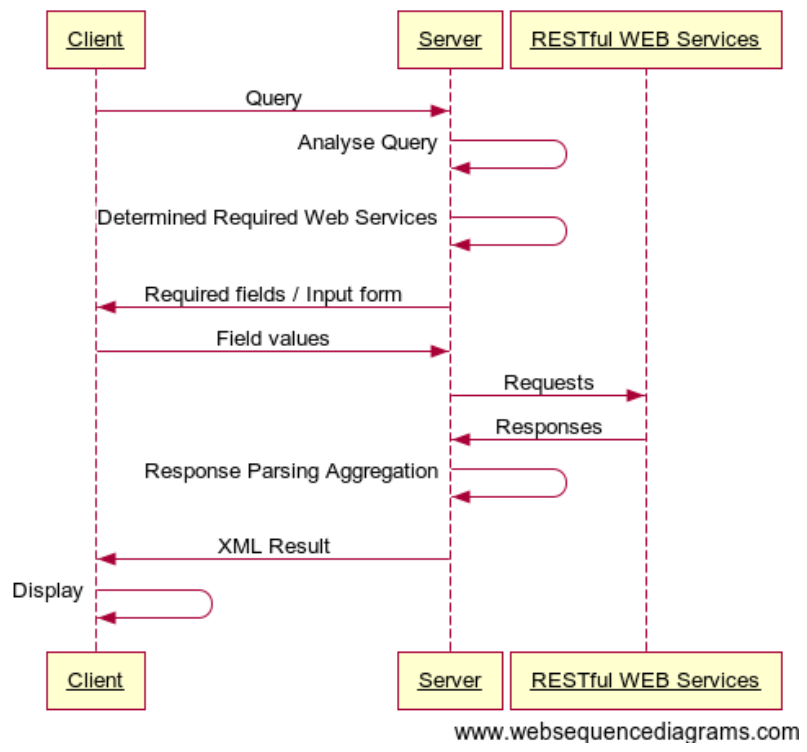

Figure 10 Interaction of SMART components

Figure 11 shows the interaction between components in the case of the mashup of two services.

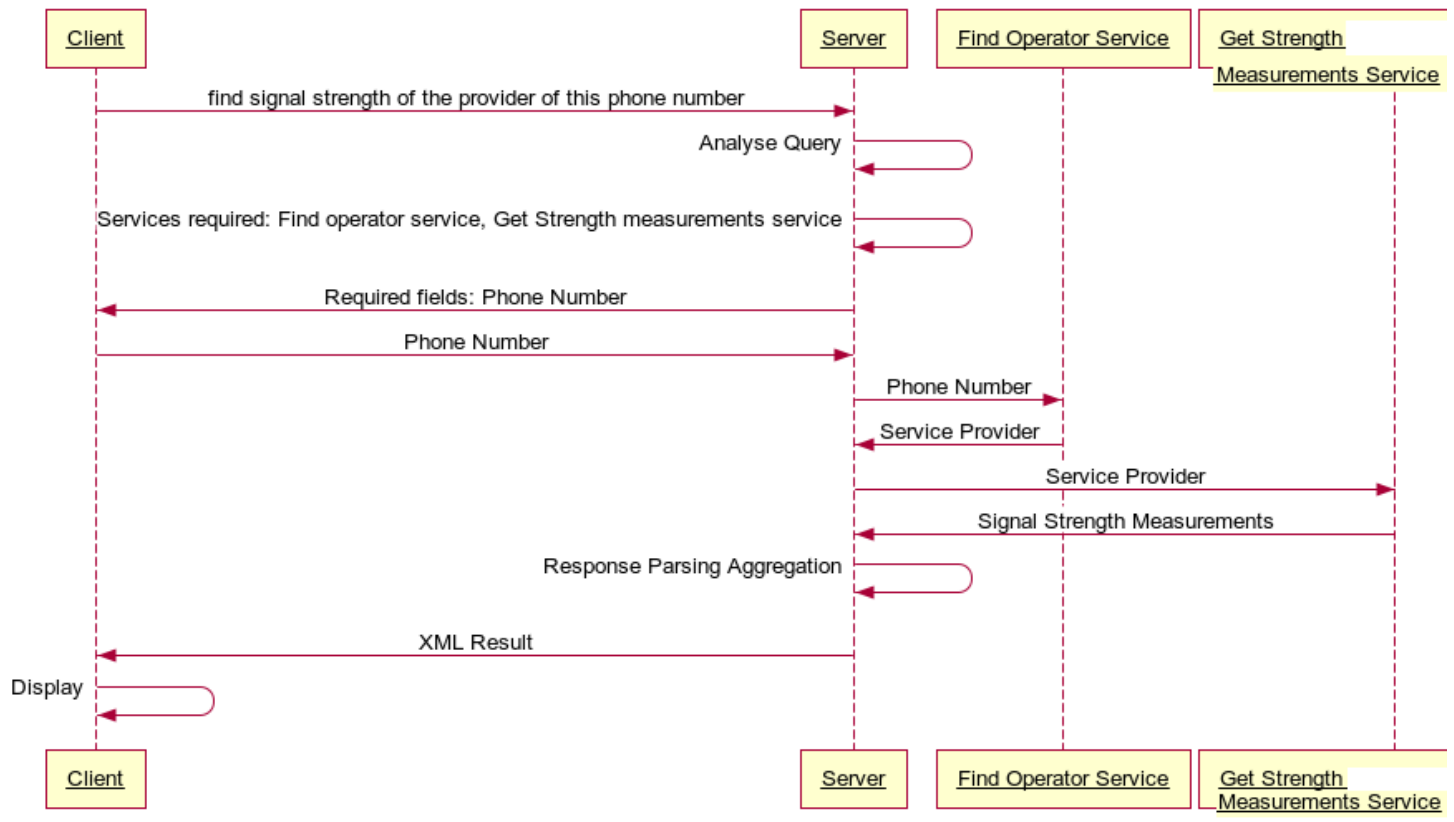

Figure 11 Mashup Example of 2 web services

In fact, the Server component, encapsulates the SMART execution engine, which does mainly analyze the request as we will see in Section 4, and search the services.owl ontology file to determine the requested web services, in order to execute the user request, as detailed in Section 5.

### 3.3. SMART display

The execution of a SMART service or a SMART mashup will be a list of elements, and if those elements encapsulate geographic coordinates, they will be also shown on a map.

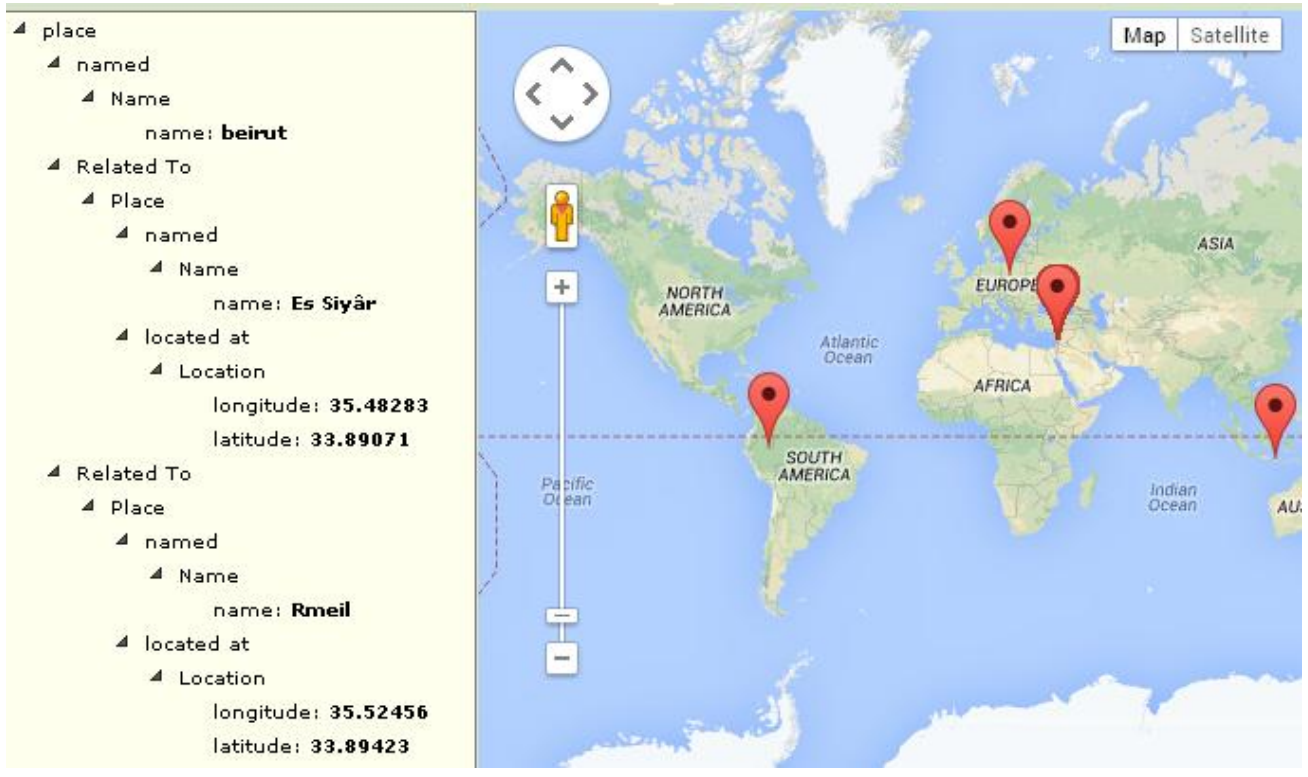

Figure 12 "find places related to this place" Output

## 4. USER QUERIES: SMARTWEB INTERFACE

In this section we describe the interaction with the user, through the web interface that we designed to be as simple as possible, very much like the Google search page, with only a TextBox where the user types his search request.

First, we will explain how the analysis of the user entry is done by taking two different examples:

- Example 1: We suppose that the user types: "find places related to **this** place"
  This request will lead us to implicitly invoke a well-known REST Web service on the internet known as the Geonames.org service, which we semantically described and added to the ontology as an individual of the SISOService class, under the name: "Geonames.org Search Service", with all the necessary information, as we can see in Figure 14. It simply waits for a place name (see Figure 13), and finds all other geographic places related to this name, showing them in a list and on a map as in Figure 12.

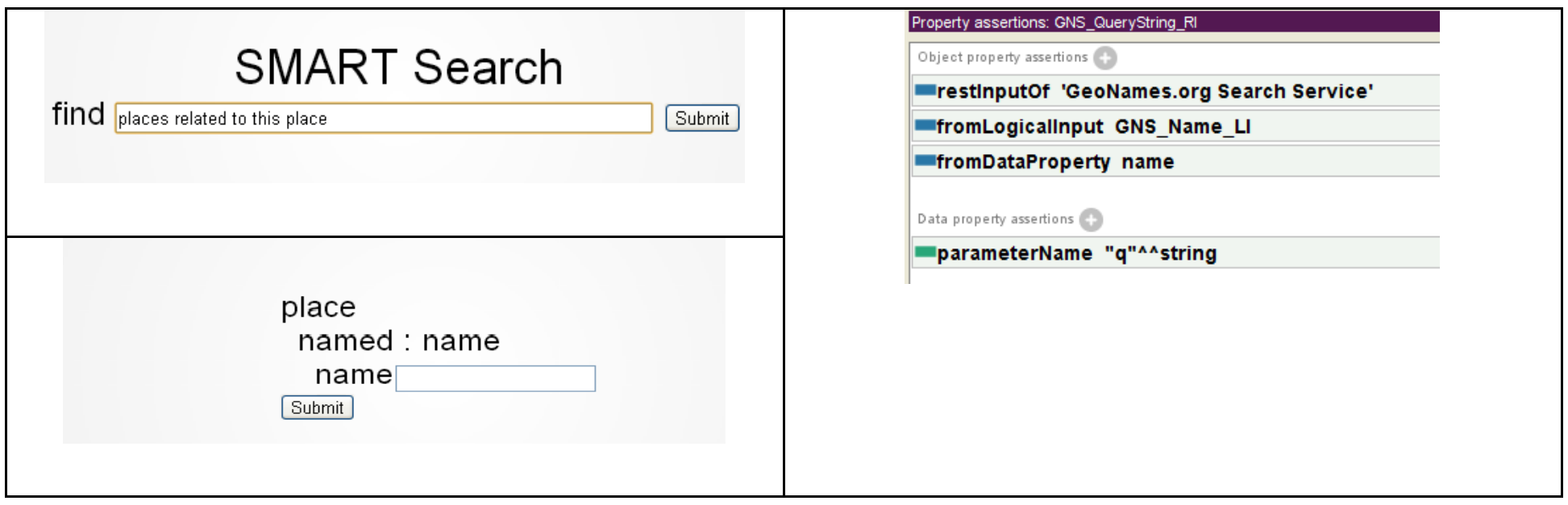

Figure 13 GeoNamesSearch service Web interface

- Example 2: We suppose that the user types: "find *the* provider of **this** phone number"

This request will lead us to invoke a REST Web service we implemented, semantically described and added to the ontology as an individual of the SISOService class, under the name: "GetOperatorService", with all the necessary information, as we can see in Figure 15. It simply waits for a phone number (see Figure 16), and returns the telecom operator attached to It as in Figure 17.

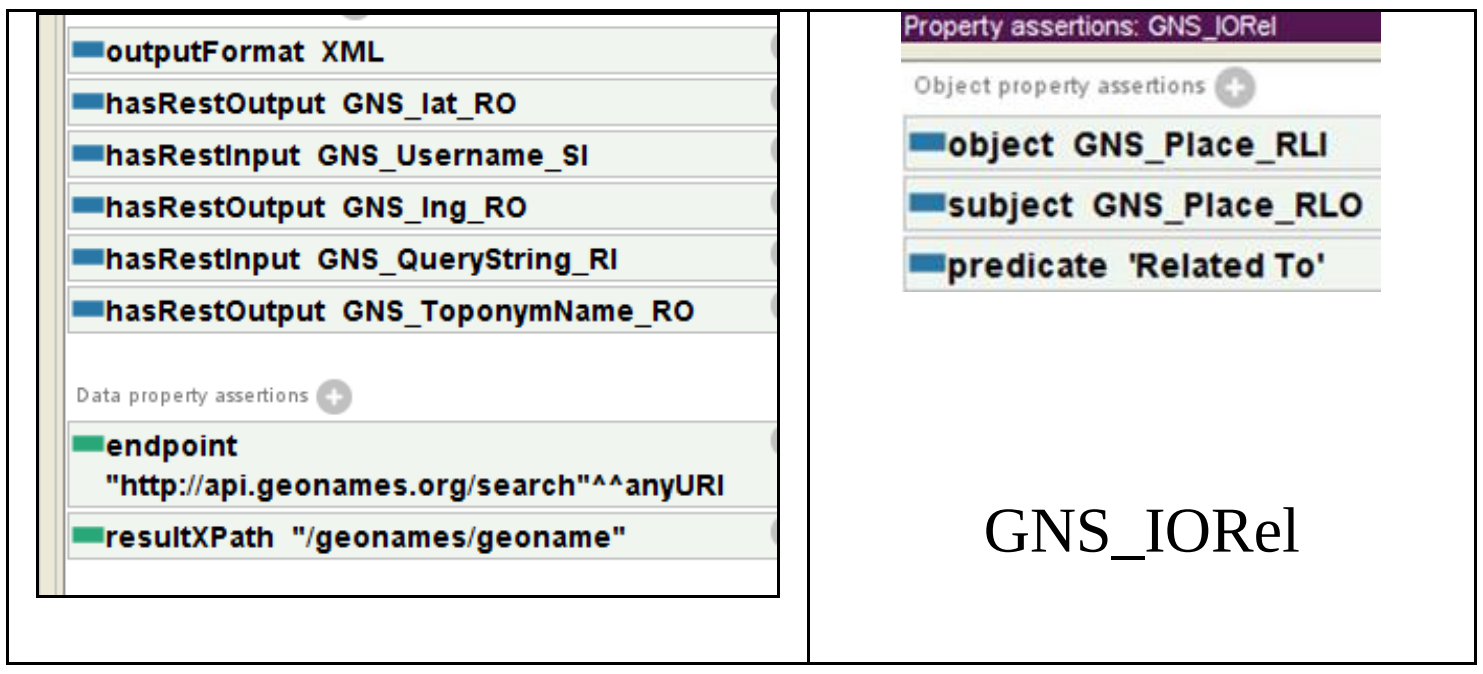

Figure 14 Geonames.org Search Service input output properties

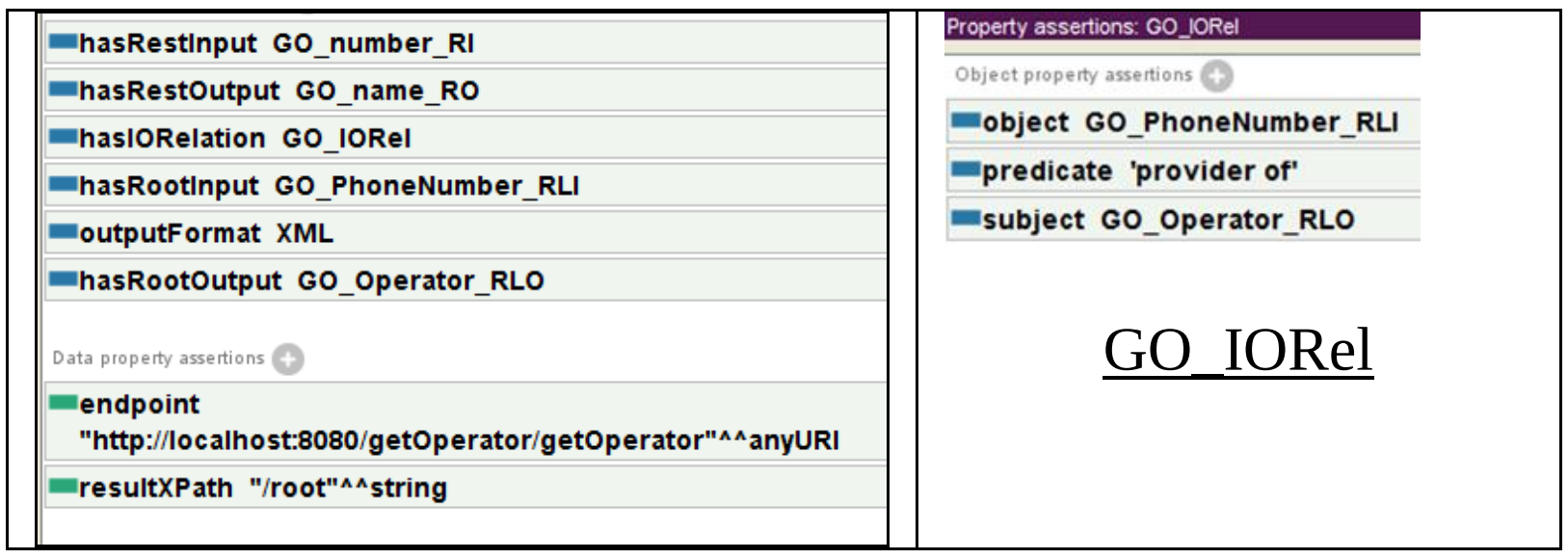

Figure 15 GetOperatorService input output Properties

One can see that any query should match the following format:

find {[<class>] <predicate>} **this** <class>

Some tokens are not mandatory, like those in italic in Example1 and Example2. The tokens between braces may repeat, while the tokens in brackets may be omitted. The token **this** is the key to an input web form shown to the user, and the keyword that follows is the class name defined at the ontology level, which represents the type of the input for the service that the SMART engine will search for, in order to execute or mash up.

This class can be as generic as needed, as long as the engine can find a service individual having this input type, as well as the properties that match up with the rdfs:range and rdfs:domain of the preceding predicate as we will explain in Section 5.

The input form will show dynamically generated input fields, corresponding to the semantic description of the input type class, in order to get all the data fields related to the class defined in the services ontology, and that represents the entry to the service to be called.

In the case of Example1, when the user submits the query he typed, he will get a web form showing one Textfield for the input of the place's name, as in Figure 13. In this case we can extract from the search keywords: the input type class "place", and the property: "related to". And since we have defined the "geonames.org Search Service" to have this input type and this IORel predicate property between its input and its output, the search for such a service in the ontology will return a reference to the corresponding individual in the ontology. This individual's endpoint DataProperty contains the URI needed to execute the call of the REST web service: http://api.geonames.org/search. Exactly like with a puzzle, the place name entered by the user fills its place in the parameter part of the service REST input, and the service is called.

In the case of example2, exactly like in example1, "Phone Number" is the input class associated to the GetOperatorService we defined, "Provider Of" is the DomainObjectProperty, which is the predicate of the GO_IORel associated to this service, and having for subject an GO_Operator_RLO, and for object a GO_PhoneNumber_RLI. The "provider Of" predicate, is defined to have a rdfs:domain of type ServiceProvider and a rdfs:range of PhoneNumber. Clicking on the submit button, will get the SMART engine to find the GetOperatorService individual and to execute or mashup it up.

Since the answer, in both cases, is a well-defined xml structure, which is also described by the service resultXPath Data property, the SMART engine can extract the specific elements from the answer.

## 5. SERVICE EXECUTION DETAILS

The ontology file services.owl contains only the ServiceThing sub-classes definitions (Parameter, Service, InputOutputRelation, etc.), and SISOService individuals (GetOperatorService, TrueCallerReverseLookup, etc…). These individuals are in fact the registered services. Object properties of these individuals will be matched up with the user query input, in order to find the REST web service to call (cf. Section 4). To make it clear, the execution of a SMART service is done in 4 main steps: The initialization step, the match-up step, the execution step, and the output building step.

The initialization step, which occurs one time at the startup of a SMART application, consists of going through the ontology file services.owl in order to load all the registered services individuals from the ontology in a singleton class: the ServiceRegistry class. This class encapsulates a HashMap that will be populated with all the services instances, indexed by their URI. For example for the GeoNamesSearch service the URI would be: http://www.semanticweb.org/ SMART/services.owl#GeoNamesSearch. This HashMap will be looked up each time a service is identified from the user query analysis: this is what we call the matching process.

The matching of a service is done first by instantiating a subQuery object that encapsulates three components: the _inputType Ontology class, an ontology property, and the _outputType ontology class. These components are simply the tokens identified from the user input text: The matching process uses the SPARQL query language, to try to find a service in the ontology respecting those three components and the relation that ties them together.
In the case of a mashup, the user query analysis will return as many subQuery objects as there are properties or predicates in the user input text.
For example: the call matchService(null, providerOf, PhoneNumber) will be executed in the case of a simple input like : "find the provider of this phone number".
And in the case of a mashup like: "find the signal strength of the provider of this phone number", the output type class of the first subquery is the entry type class for the following subquery.
The SubQuery1 would be: (PhoneNumber, providerOf, null) and the SubQuery2 would be: (null, signalStrengthMesurementOf, null).
In fact matchService(null, signalStrengthMesurementOf, null) is equivalent to matchService(ServiceProvider, signalStrengthMesurementOf, SignalStrengthMesurement), since any information could be inferred into the most generic type, if a specific type is not provided by the request tokens.

The result of this matching process is a service from the ontology and its URI is used to lookup the HashMap, and return a reference to the corresponding service individual instance in this HashMap.

Once the service reference is found, its execute method is called with the list of the input individuals. In our case, it is always one Logical root input individual, with logical and REST sub-elements (see Section 2.1). This is the execution step that begins by identifying the service input which can be filled in by the dynamically generated user form, or from the output of another service call (in the case of a mashup). Next, a call to the method buildURL builds up the REST call, by concatenating entry values in a puzzle like way, into the endpoint URL DataProperty of the service adding whatever extra constant parameters which names and values are also stored in the ontology (Ex: apiKey=21o2iu34oiu1234). Finally, having a valid URL, the REST web service is called.

The last step is the output building step: The response to the service call may be in XML or JSON format, in the latter it is converted into XML. A parse() method is then applied to the XML result in order to build up the output DomainThing individuals into memory, for the duration of a user session. These individuals are never persisted into the services.owl file. They are graph-like memory objects constructed by triplets (subject, predicate, object) as follows:

- For each result, consult the resultXPath of the service.
- For each root individual (in fact we have only one LogicalRootOutput) consult the rootOutputXPath, create a new individual, and create sub-individuals (with as many levels as necessary ) and tie them using the values of the properties fromObjectProperty of the corresponding LogicalOutputParameter.
- In order to fill out the DatatypeProperties, for each RestOuputParameter evaluate the restOutputXPath expression in order to access the literal value of the needed DatatypeProperty.

Figure 17 shows the leveled output of the getOperator service.

## 6. STEPS TO ADD A NEW SMART SERVICE TO THE ONTOLOGY

In order to be able to mashup or even simply call a REST web service, the service should be well defined (as we will expose next) and added to the ontology services.owl file. At the application level, a simple restart is needed in order to reload the ontology file and rebuild the services dictionary (cf Section 5). These two tasks are time-consuming, so the first call of the application will always be slower than the subsequent calls.

The most straightforward way to explain these steps is by using the example 2 from Section 4, where the user is searching for the mobile operator (provider) of a certain cellular number, by invoking a REST web service already up and running, the GetOperatorService. By clicking on the submit button, a dynamically generated form, is generated in order to input the number (see Figure 16):

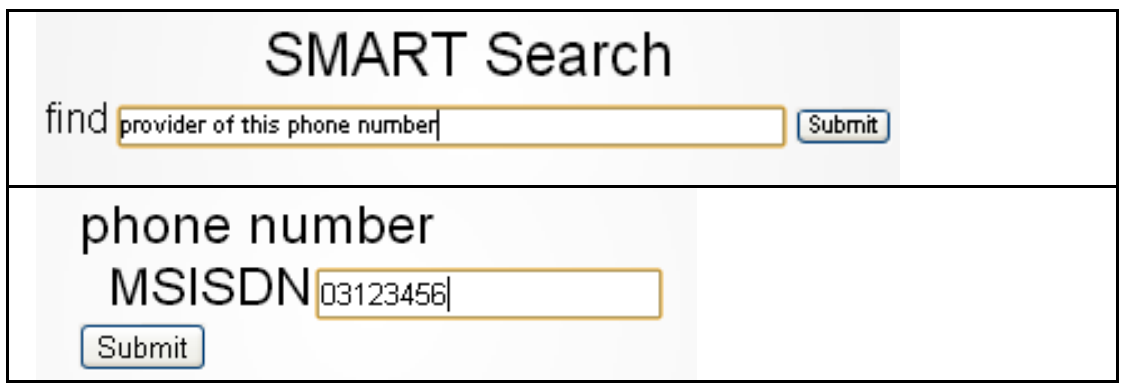

Figure 16 GetOperator Service Web interface

The output will be as shown in Figure 17:

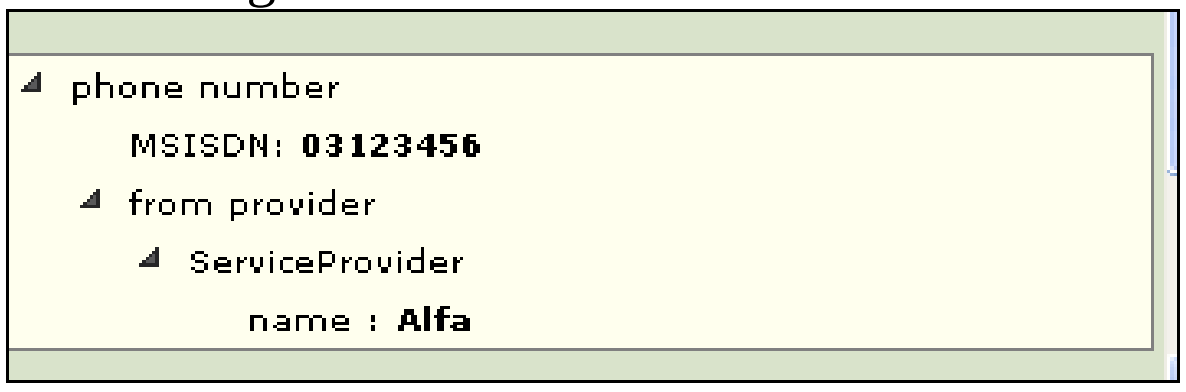

Figure 17 GetOperator Service Output

This particular service has been defined at the ontology level as in Figure 15.

Table 2 GetOperatorService Object Properties

| | |
|---|---|
| 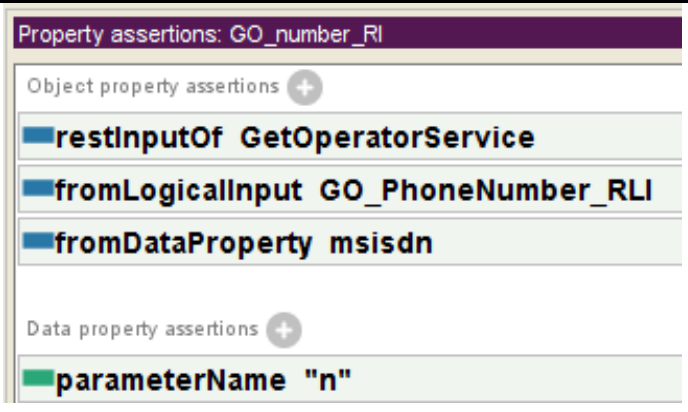<br>GO_number_RI:<br>VariableRestInputParameter | 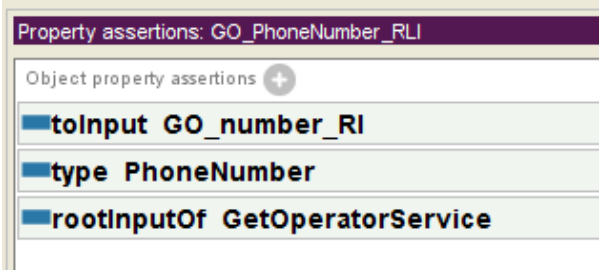<br>GO_PhoneNumber_RLI:<br>RootInputParameter |
| 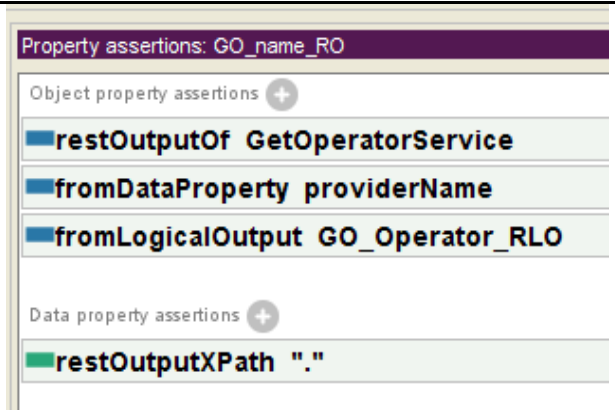<br>GO_name_RO:<br>RestOutputParameter | 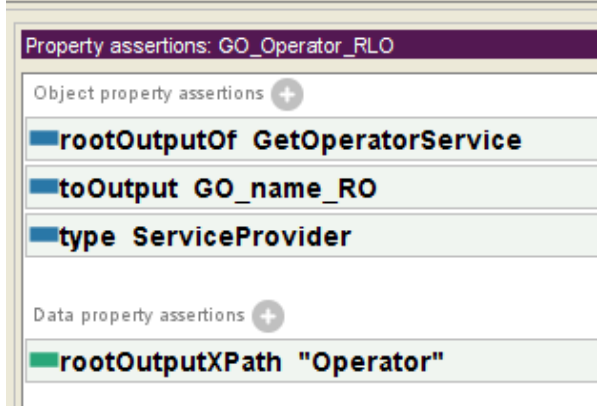<br>GO_Operator_RLO:<br>RootOutputParameter |
| 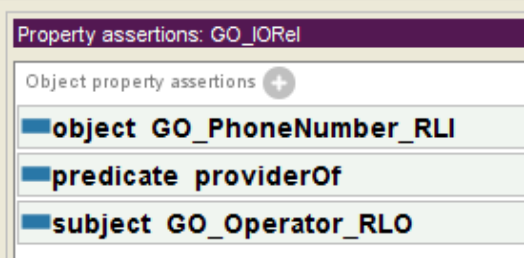<br>GO_IORel:<br>InputOutputRelation | 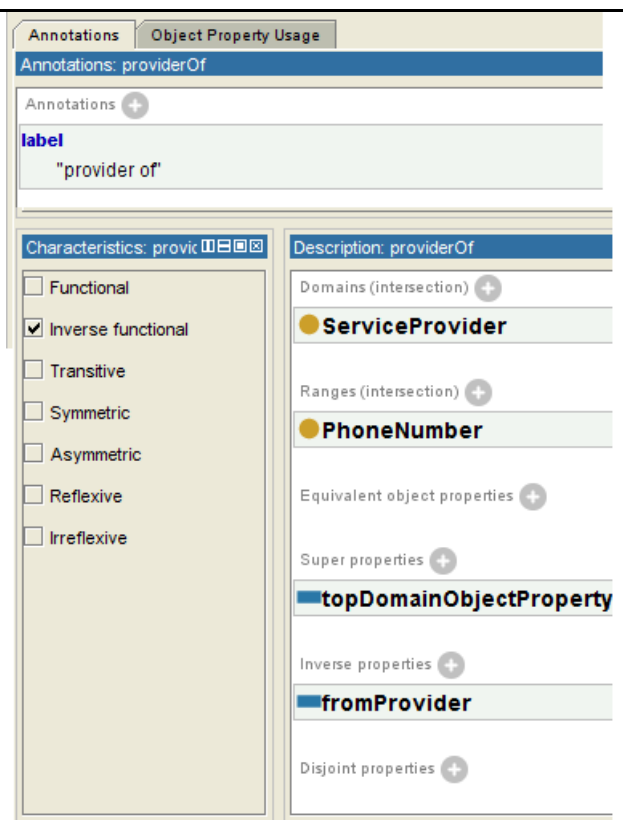<br>ProviderOf:<br>Predicate |

The steps to achieve this task are very simple (see **Table 2**):

- Add an individual of the SISOService class (We are considering the case of web services with single Root entries, and outputs) : GetOperatorService

- Define the input parameters: hasRestInput GO_number_RI, and hasRootInput GO_PhoneNumber_RLI
- Define the ouput parameters: hasRestOuput GO_name_RO and hasRootOutput GO_Operator_RLO
- Define the relation between inputs and outputs: hasIORelation GO_IORel

All the underlined keywords are Object Properties of the GetOperatorService individual.

## 7. CONCLUSIONS AND FUTURE WORK

In this paper, we have described SMART, our user-oriented semantic Web platform for mashing up REST web services.

The creation of useful mashups such as: find me something and show it on a map, find me signal strength, a person, a book, a measurement, becomes as simple as typing a search request.

SMART offers contributions at two main levels: The first contribution is materialized by an ontology for describing REST Web services. The second contribution is the web framework where any user can simply mashup REST Web services by using the keywords corresponding to his search request.

The benefits of semantically describing services arise from the ontology reasoner, which can infer the services to call, by matching up the user keywords for the predicate and the input output classes, even if those keywords were not exactly the ones used for the semantic description of the service. If a service cannot be found, a more generic service may be returned.

Another benefit is tied to the expressiveness of the semantic description of the classes and the properties, since we can define equivalent properties and classes having different labels, so the user can use these synonyms in his request.

At the level of describing the service properties, the use of SWRL gave us the possibility to skip the definition of certain information such as the RootInput and the IORelation which can be thus inferred.

The SMART platform implemented as a proof-of-concept, is publically deployed at the URL: github.com/anthonyhseb/SMART. At a short term, a web interface will be developed in order to ease the task of adding new web services descriptions to the ontology file. At a longer term, the ontology should cover the description of services with multiple entries. Finally, one of the future tasks that will greatly benefit this work is the integration with yahoo pipes.

## ACKNOWLEDGEMENTS

The authors would like to thank Anthony Hessab and Mohammad Mbadder at the Saint-Joseph University for their contribution and related work that benefited greatly this paper.

**Authors**

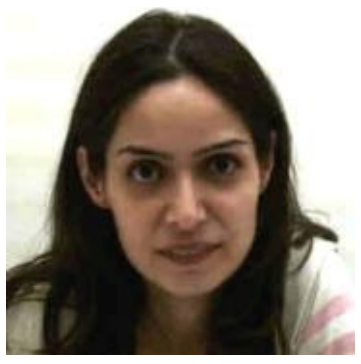

Rima Kilany is an Associate Professor at the “Ecole Supérieure des Ingénieurs de Beyrouth” of Saint-Jospeh University, Lebanon.

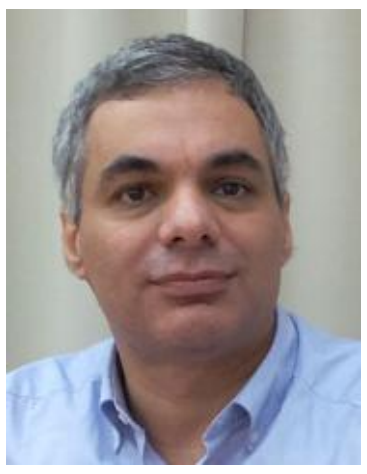

Maroun Chamoun is the Director of the CIMTI: « Centre d'Informatique, de Modélisation et de Technique de l'Information » and an Associate Professor at the « Ecole Supérieure des Ingénieurs de Beyrouth » of Saint Jospeh University, Lebanon.